\documentclass[a4paper,fleqn]{cas-dc}

\usepackage[numbers]{natbib}
\usepackage{float}
\def\tsc#1{\csdef{#1}{\textsc{\lowercase{#1}}\xspace}}
\tsc{WGM}
\tsc{QE}

\begin{document}
\let\WriteBookmarks\relax
\def\floatpagepagefraction{1}
\def\textpagefraction{.001}

\shorttitle{Development of millikelvin scintillation photodetectors with CUPID}    

\shortauthors{Nicola Manenti}  

\title [mode = title]{Development of millikelvin scintillation photodetectors with CUPID}  

%

\author[1,2]{Nicola Manenti}[orcid=0009-0008-9942-5127]

\cormark[1]

\ead{nicolamanenti01@universitadipavia.it}

\credit{}

\affiliation[1]{organization={University of Pavia},
            addressline={Via Bassi 6}, 
            city={Pavia -},
            citysep={}, 
            postcode={27100}, 
            country={Italy}}
\affiliation[2]{organization={INFN Pavia Section},
            addressline={Via Bassi 6}, 
            city={Pavia -},
            citysep={}, 
            postcode={27100}, 
            country={Italy}}

\author{on behalf of the CUPID Collaboration}

\credit{}

\cortext[1]{Corresponding author}


\begin{abstract}
CUPID is a next-generation tonne-scale cryogenic experiment that will search for the neutrinoless double beta decay (\(0\nu\beta\beta\)) of \(^{100}\)Mo. It will be located at the Gran Sasso National Laboratory (Italy) and will employ the cryogenic calorimetric technique to reconstruct the energy deposited in Li\(_{2}\)MoO\(_{4}\) crystals, in which the isotope of interest is embedded. Preliminary measurements have been performed with the CUPID VSTT (Vertical Slice Test Tower), a prototype tower designed to test the detectors’ response and the apparatus’ mechanical and thermal stability. In addition, a key objective was to demonstrate the improved performance of germanium-based cryogenic light detectors. In CUPID, light detection is used for both particle identification and pile-up rejection. Li\(_{2}\)MoO\(_{4}\) is a scintillating crystal, enabling particle identification by exploiting the dependence of light yield on particle type. This requires collecting the Li\(_{2}\)MoO\(_{4}\) scintillation light, peaking at \(\sim\)600 nm, with a \(\sim\)300 \(\mu\)m-thick high-purity germanium wafer placed next to the crystal. The wafer is also equipped with concentric Al electrodes biased at a few tens of volts. This bias improves the signal-to-noise ratio by an order of magnitude thanks to the Neganov-Trofimov-Luke effect. This work presents preliminary results on the performance of the VSTT light detectors and of its Optical Injection System.
\end{abstract}

\begin{keywords}
 Cryogenic bolometers\sep Neutrinoless double beta decay\sep Neganov-Trofimov-Luke effect\sep
\end{keywords}
\maketitle
\section{Introduction}\label{s1}
Neutrinoless double beta decay (\(0\nu\beta\beta\)) is a hypothetical process whose observation would establish lepton-number violation and demonstrate that neutrinos have a Majorana mass component. Neglecting nuclear recoil, the summed kinetic energy of the two emitted electrons equals a  Q\(_{\beta\beta}\) on the order of a few MeV, yielding a monoenergetic peak in the summed-energy spectrum Experimental searches have established isotope-dependent lower limits on the half-life at the level of 10\(^{24}\)–10\(^{26}\) yr. For rare event searches, higher statistics can result from a longer exposure time or a larger experimental volume containing a bigger sample of candidate isotopes, while the energy resolution affects the final sensitivity. The available detection techniques range from monolithic detectors, such as scintillators or Time Projection Chambers, to granular ones, like semiconductors and bolometers.
Bolometers are small detectors, typically a few centimetres per side, operated at cryogenic temperatures of \(\sim\)10 mK. They are capable of detecting the small temperature increase produced by an energy deposition, with a typical sensitivity of approximately hundreds of \(\mu\)K/MeV. In the context of \(0\nu\beta\beta\) searches, bolometers offer the advantage of combining source and detector in a single device. They are made up of three main components: an absorber, a thermal sensor, and a thermal link to the heat bath. The absorber is a crystal containing the isotope of interest and therefore acts simultaneously as the source and the detection medium. The thermal sensor, for example a thermistor, has a temperature-dependent response. Finally, the thermal link connects the absorber to the environment and allows the detector to return to its nominal operating temperature after an event. Bolometers have an excellent energy resolution, typically at the level of a few parts per thousand. However, a large number of them are needed to achieve the same integrated active mass that can be obtained with a monolithic detector.
\section{Experimental Details}\label{s2}
\begin{figure}
  \centering
    \includegraphics[width=0.7\columnwidth]{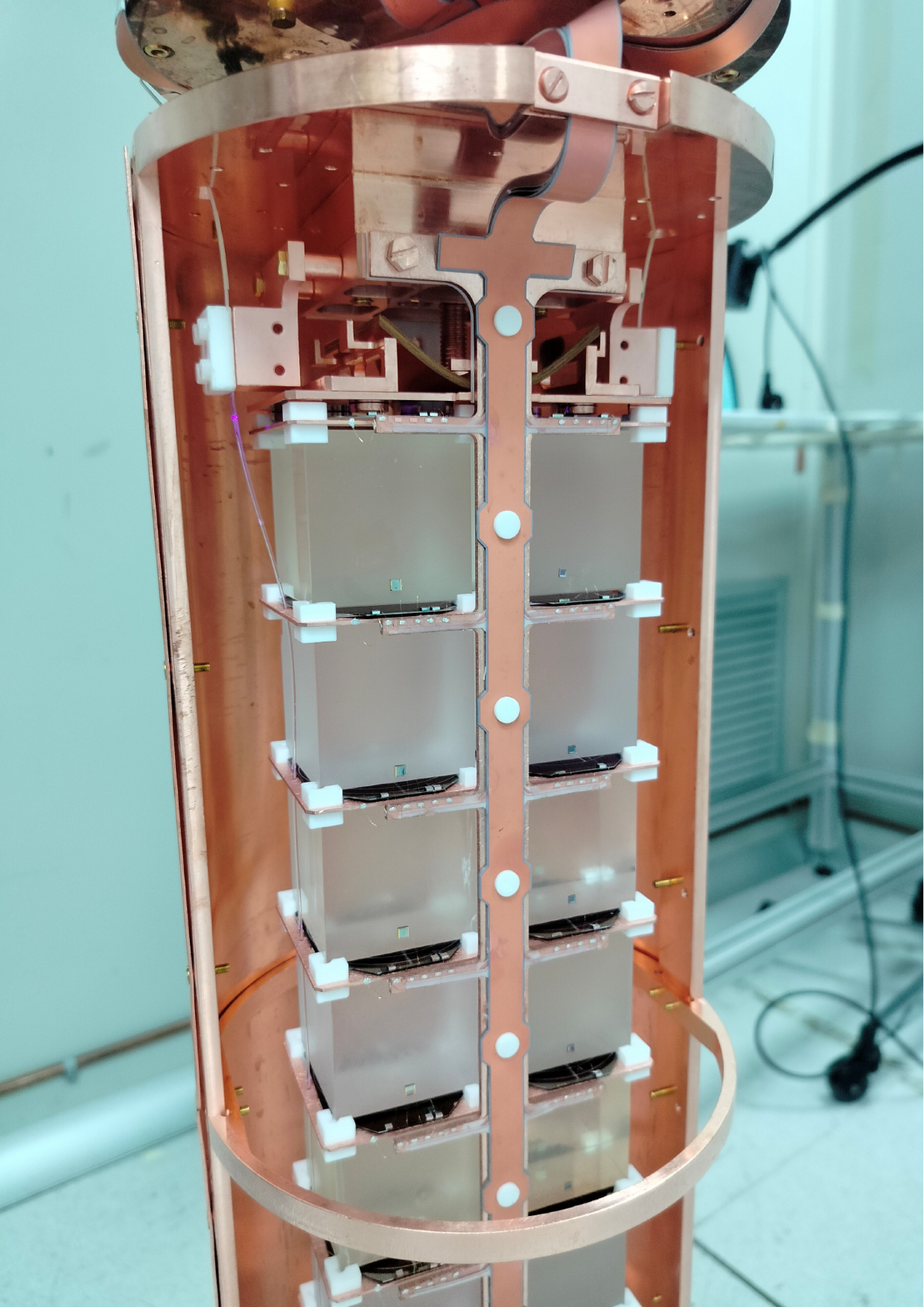}
    \caption{Upper section of the VSTT prototype tower. The complete tower comprises 28 Li\(_2\)MoO\(_4\) crystals arranged in two columns and interleaved with 30 germanium light detectors.}\label{FIG:1}
\end{figure}

CUPID \citep{CUPID:2025avs} is a next-generation cryogenic experiment for the \(0\nu\beta\beta\) decay of \(^{100}\)Mo, one of the few candidate isotopes that can undergo this process  (\(Q_{\beta\beta}\) = 3034 keV). Currently in its R\(\&\)D phase, it will deploy about 1600 Li\(_{2}\)MoO\(_{4}\) (LMO) crystals (45×45×45 mm\(^{3}\)), enriched in \(^{100}\)Mo, with an expected livetime of ten years. These crystals will be operated as bolometers (fig.\ref{FIG:1}) below 15 mK and split among 57 towers. Bolometers are capable of collecting the energy released following a decay event.  Provided that the decay is contained inside the absorber crystal, the energy deposited is converted into crystal lattice vibrations (phonons), thus increasing the absorber temperature. The readout is performed with a thermometer: specifically, a Neutron Transmutation Doped (NTD) Ge sensor, whose resistance decreases as the temperature increases. These detectors require cryogenic temperatures and use a \(^{3}\)He-\(^{4}\)He dilution refrigerator to maintain a stable working temperature around 10 mK. Furthermore, they are extremely slow, with an overall duration of several hundreds of ms, and require very low background rates. The reconstructed energy spectra are used to search for a signal and, in its absence, to set a lower limit on the \(^{100}\)Mo \(0\nu\beta\beta\) half-life. Some challenges in employing this technique for the study of \(0\nu\beta\beta\) have already been encountered in the CUORE experiment, whose facility will be repurposed for CUPID use. Unlike CUPID, CUORE  \citep{CUORE:2024ikf} searches for the \(0\nu\beta\beta\) decay of \(^{130}\)Te. Thanks to its data-driven background model \citep{CUORE:2024fak}, which resulted from the analysis of an experimental exposure of 1038.4 kg\(\cdot\)yr, CUORE works as a reliable benchmark when delineating next-generation experimental challenges. Hence, with this configuration only, there are two known limiting factors in the achievable sensitivity: the rate associated with background sources, mainly energy-degraded alpha particles in the MeV energy region, and pile-up events from the always present \(2\nu\beta\beta\) decays.
\subsection{Alpha background and light yield-based discrimination}
\begin{figure*}
	\centering
	\includegraphics[width=.9\textwidth]{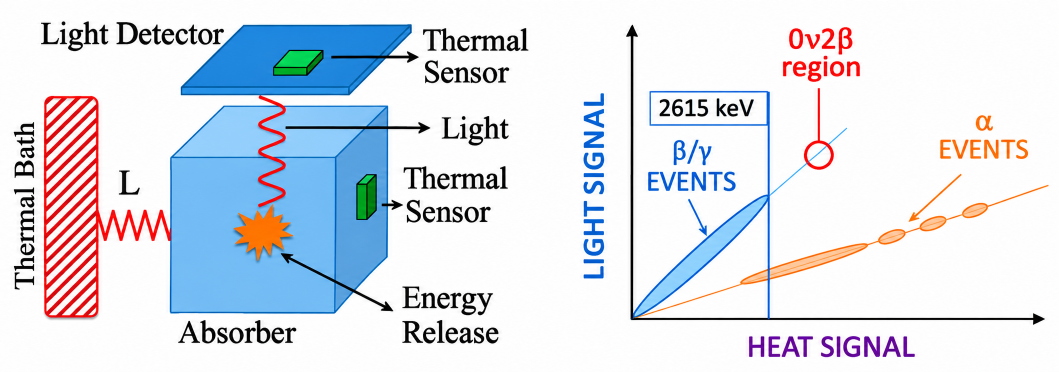}
    \caption{Left: schematic operating principle of a scintillating bolometer with separate heat and light readout.
    The heat signal is read from the crystal absorber containing the \(0\nu\beta\beta\) candidate isotope, while the light signal is detected on a second absorber sensitive to the scintillation light. The temperature rise in each absorber is measured by their respective thermal sensors. Right: Concept of alpha particle rejection by simultaneously measuring the light and heat signals and exploiting the lower light yield of alpha interactions relative to \(\beta/\gamma\) ones.}\label{FIG:2}
\end{figure*}

Alphas are the limiting background factor for CUORE sensitivity. They are in general produced on the crystal surface or in the materials surrounding the detectors. The current CUPID baseline plans to perform a dual readout acquisition to discriminate between alphas and betas/gammas. More specifically, the plan is to use Li\(_{2}\)MoO\(_{4}\), which are scintillating, and to collect the signal in both heat and light channels. The detection of the scintillation light release from LMO crystals, which peaks around 600 nm, is performed by introducing additional light detectors (LDs), fig.\ref{FIG:2}. Similar to LMO crystals, they are Ge-wafers working as bolometers and read out through NTD sensors to reach ~ 0.3 keV/MeV of detection light-yield for betas/gammas in the LMOs, which has been proven to be feasible \citep{CUPID:2025wbt}, while alpha interactions produce approximately five times less light at the same deposited energy.
CUPID will record the detectors' signal as a continuous data stream with a sampling frequency of a few kHz, which slightly differs between LMOs and LDs. LDs will require a better time resolution to meet the CUPID baseline requirements, thus requiring a sampling rate of 10 kHz, with respect to 1 kHz that will be used for LMOs. Each LMO is placed between two LDs, one on the top and another one on the bottom. By collecting the continuous detectors' response, when an event occurs in a specific LMO, it is then possible, with an offline analysis, to probe the scintillation light amount detected in its top or bottom LD. This makes light yield-based discrimination possible. The main advantage of amplifying the scintillation light coming from the LMO is that even in the case of sub-optimal LD response, it is possible to boost the signal-to-noise ratio by increasing the voltage bias of the electrodes, since the noise is not amplified in the same way as the scintillation signal. 
\subsection{\(2\nu\beta\beta\) pile-up and Neganov-Trofimov-Luke amplification}
\begin{figure}
  \centering
    \includegraphics[width=1\columnwidth]{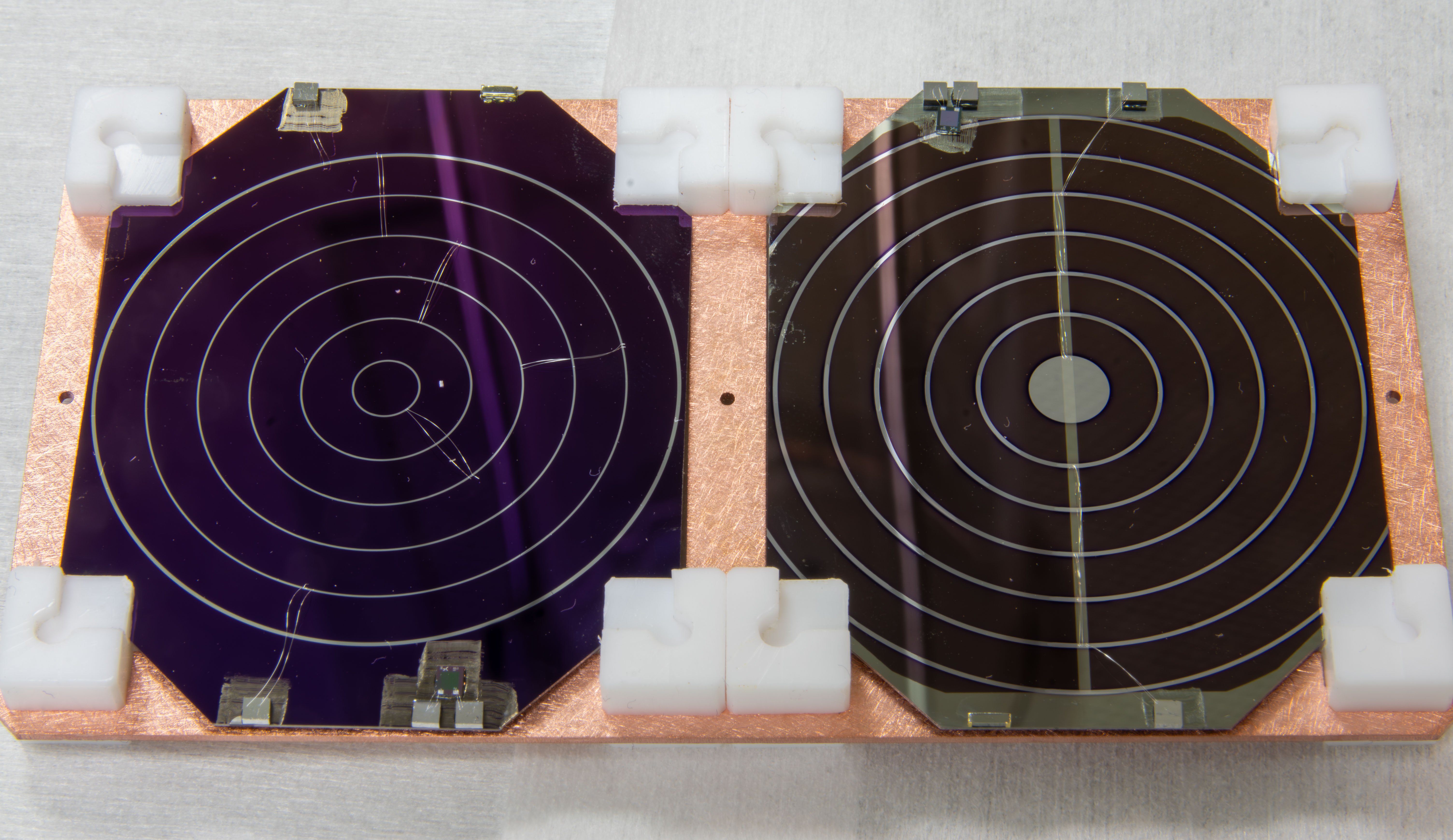}
    \caption{Close-up to a couple of LD with the concentric electrodes for the NTL amplification. }\label{FIG:3}
\end{figure}

\begin{figure}[]
  \centering
    \includegraphics[width=1\columnwidth]{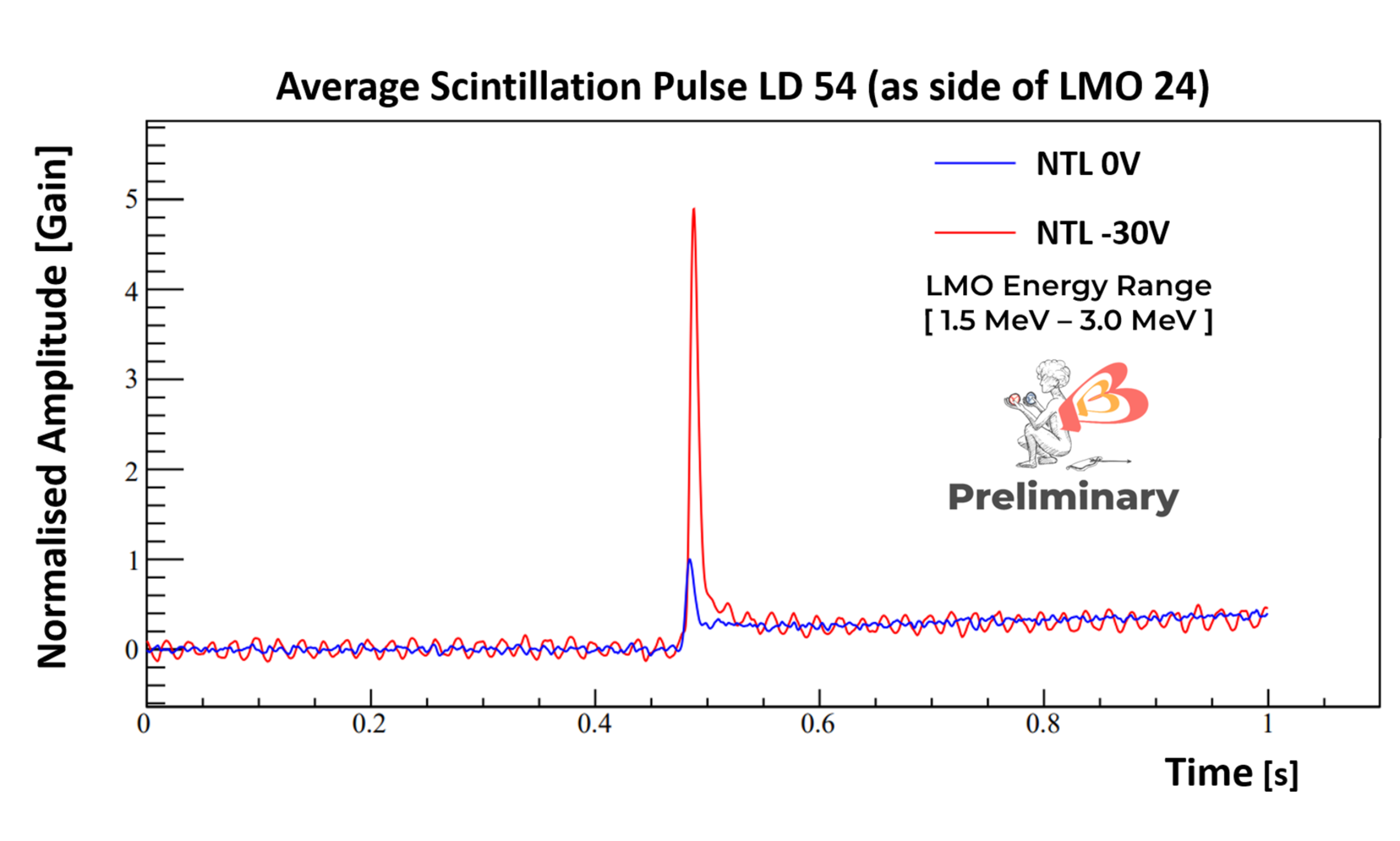}
    \caption{Average pulses recorded by LD-54 at electrode biases of 0 V and -30 V, for events whose reconstructed energy by the analysis in the adjacent LMO crystal lies between 1.5 MeV and 3.0 MeV. Both traces are normalised to the average pulse amplitude at 0 V.}\label{FIG:4}
\end{figure}

\begin{figure}
  \centering
    \includegraphics[width=1\columnwidth]{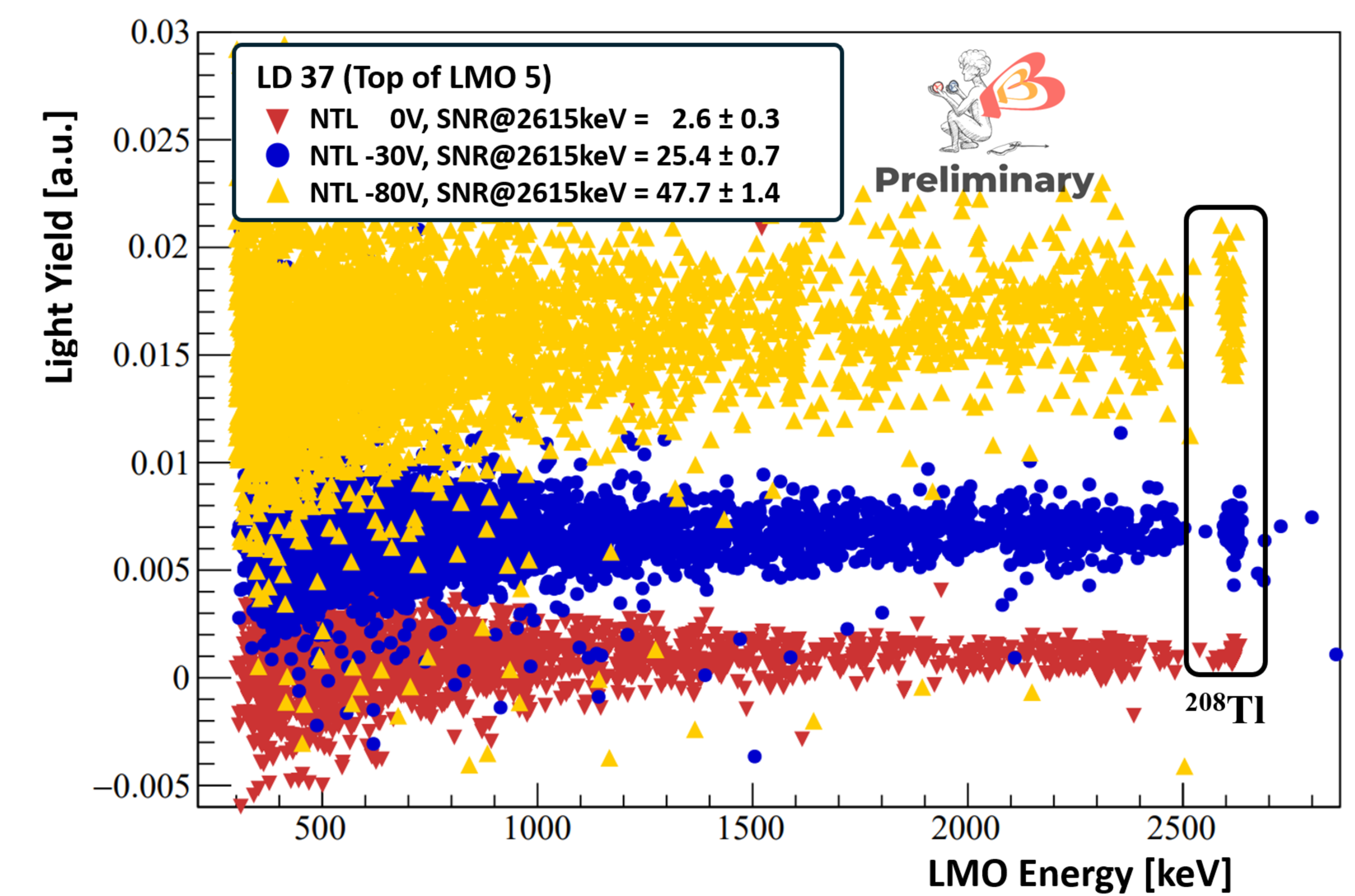}
    \caption{NTL amplification of the scintillation-light signal recorded by VSTT light detector LD-37 at electrode biases of 0, -30, and -80 V. The light-detector response per unit energy deposited in the associated LMO crystal is shown in arbitrary units. A common offset is subtracted from all datasets so that the median of the 0 V dataset is centred at zero. The SNR values evaluated at the 2615 keV \(^{208}\)Tl peak are reported in the legend.}\label{FIG:5}
\end{figure}

Pile-up events are produced from random coincidences of more than one \(^{100}\)Mo \(2\nu\beta\beta\) event occurring in the same crystal, whose half-life is \(\sim\)7.1×10\(^{18}\) yr, if spaced less than a few hundred ms, which is the typical response time of a bolometer. To reject them, good pulse shape discrimination is fundamental. The baseline is to equip the Ge wafers, fig.\ref{FIG:3}, with electrodes and then utilise the Neganov-Trofimov-Luke (NTL) amplification effect to achieve a signal gain of a factor \(\sim\)10 with a bias of \(\sim\)100 V, thereby improving the pulse shape discrimination. Scintillation photons absorbed in the Ge wafer generate electron–hole pairs. Their drift in the applied electric field produces additional phonons, increasing the thermal signal measured by the NTD thermistor. The NTL effect \citep{Armatol_2026} increases the generated amount of phonons in a semiconductor by inducing, with an external electrical field, the drift of the aforementioned charges. In this regime, the produced number of phonons along the drift is proportional to the applied voltage, and the reconstructed signal increases linearly as:
$$E_{\mathrm{th}}=E_0\left(1+\eta\frac{e|V|}{\epsilon}\right), $$
where \(E_{\mathrm{th}}\) and \(E_0\) are respectively the total thermal energy measured and the energy absorbed in the Ge wafer, \(\eta\) is a factor that accounts for incomplete amplification, \(e\) is the electron charge, \(V\) is the magnitude of the potential difference between electrode sets and \(\epsilon\) is the average pair-creation energy appropriate to the incident radiation. The overall effect is an increase in the reconstructed signal amplitude from the thermistor response (fig.\ref{FIG:4}) and improved resolution. With the most recent R\(\&\)D tower, among other key points, we focused on the signal-to-noise performance for NTL-assisted light detectors. To provide a clear reference for the computation, we selected scintillation events on the LDs coming from the 2615 keV gamma peak of \(^{208}\)Tl, fig.\ref{FIG:5}. Then we scanned different voltage biases for the NTL and observed that, when biasing the electrodes with a bias of 80 V, there was an improvement of a factor of about 20. Despite the response enhancement, NTL-based detectors suffer from a major drawback. 
Some of the charge carriers generated by energy depositions in the Ge wafer may become trapped at defects or impurities. The resulting accumulation of space charge can distort the electric field and reduce the effective NTL gain. The detector response can be restored through regeneration, which consists of setting the electrode bias to zero and illuminating the wafer with high-intensity light to neutralise the accumulated space charge.
\section{Optical Injection System}
\begin{figure}
  \centering
    \includegraphics[width=1\columnwidth]{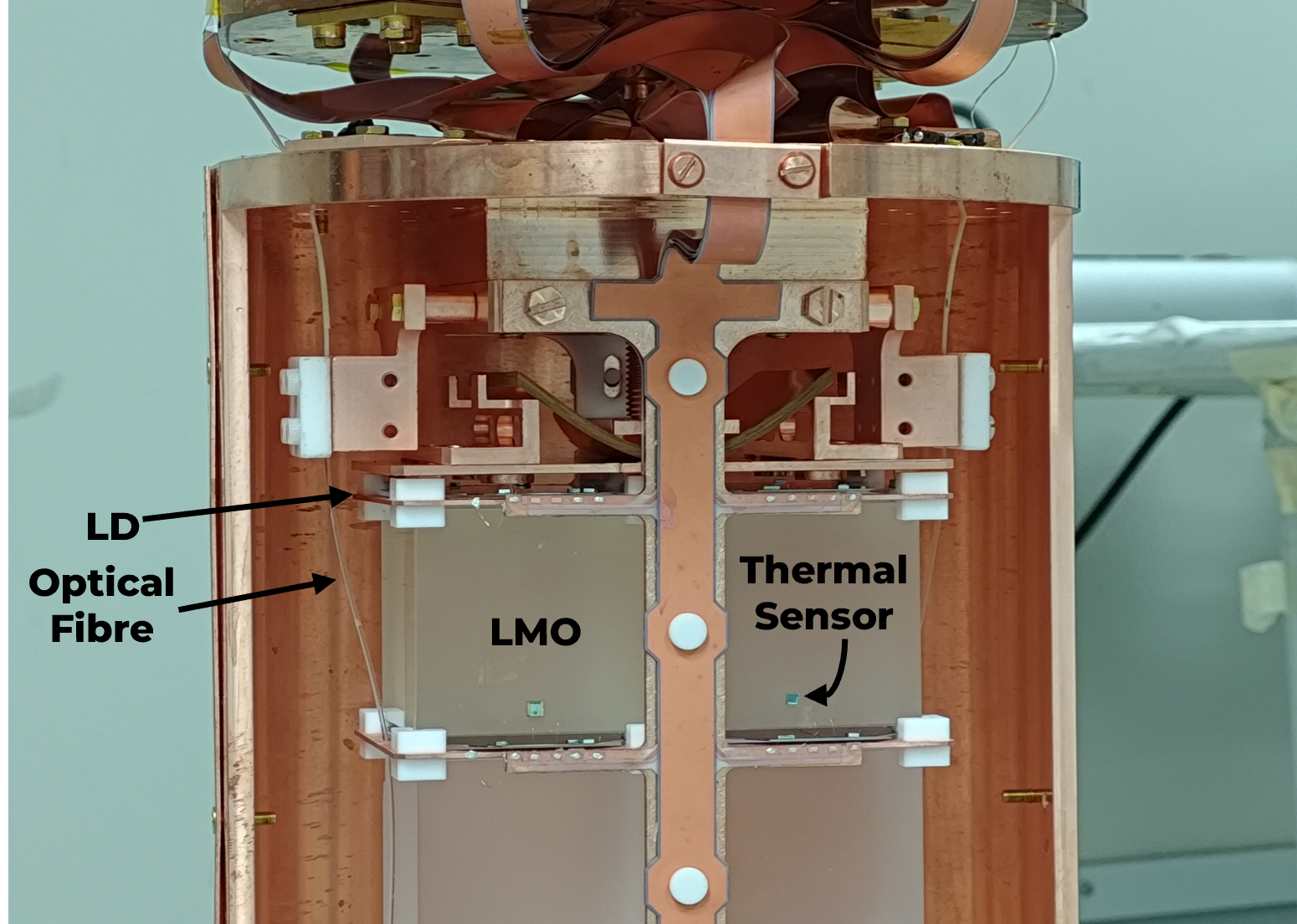}
    \caption{Close-up of the upper part of the VSTT prototype with the two optical fibres that run along the tower. On each side, a T-shaped copper-PTFE piece connects the transport fibre that travels across the different stages of the cryostat and the light-diffusing fibre which actually faces the detectors.}\label{FIG:6}
\end{figure}

To overcome this obstacle, the CUPID experiment has been developing an apparatus, the Optical Injection System (OIS), capable of producing light pulses of a given wavelength. More specifically, light pulses are generated outside the cryostat, then transported inside the inner vessel via optical fibres, and finally injected into a group of LDs. The LD regeneration \citep{Isaila_2012} is the main task of the CUPID Optical Injection System, but it does not impose strict restrictions on which light source to deploy or how it has to be driven. In the baseline design of CUPID, the OIS will inject light pulses with a wavelength similar to the scintillation light of the LMO crystals into the LDs at an adjustable rate. Nevertheless, the system is also capable of simulating pile-up events on the detector using a pair of close pulses, or of performing additional functions such as absolute calibration through a stable monochromatic source, response monitoring, and stabilisation of the LDs.
In order to verify the achievable performance, a single CUPID tower prototype, the Vertical Slice Test Tower (VSTT, fig.\ref{FIG:1}), was tested with its own OIS. We adopted a single-light-source setup with two optical fibres running along the tower, one for each side, as shown in fig.\ref{FIG:6}. The system consists of two parts: inside the cryostat, the injecting lines, and outside, at room temperature, the electronics and the light source. Fig.\ref{FIG:7} shows the whole schematic of the apparatus.
\subsection{OIS: VSTT setup}
\begin{figure}
  \centering
    \includegraphics[width=1.\columnwidth]{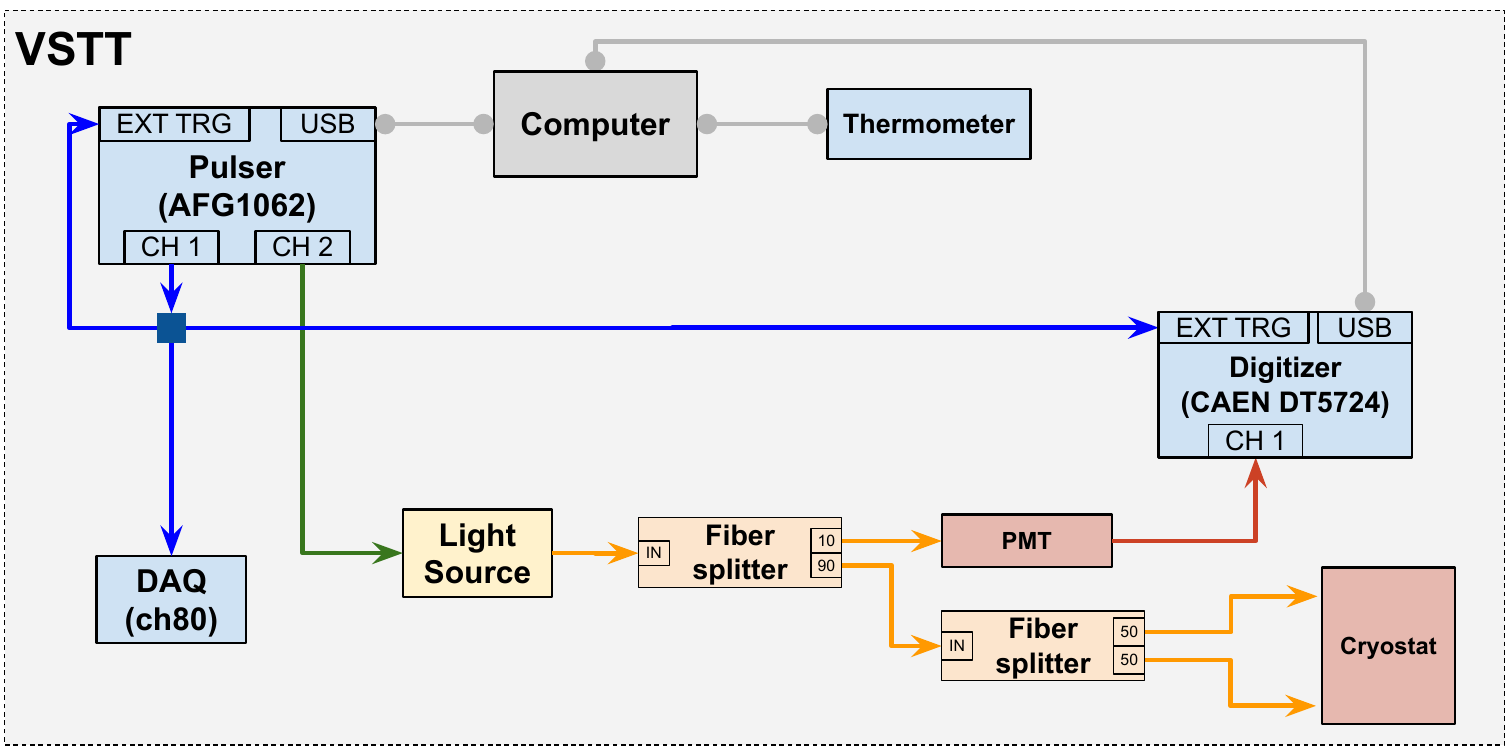}
    \caption{Schematic of the OIS setup deployed for VSTT. The pulse generator and monitoring instrumentation, PT1000 thermometer and digitiser, are all connected via USB cable, here in grey, to the computer where the OIS software is running. The output from channel 1 of the pulser, blue lines, works as a logic signal. It is split and used both as a gate signal for the outer channel of the pulser, as well as a trigger signal for the data acquisition with the digitiser and as a flag signal for the CUPID DAQ. The flag is digitised independently as a logic signal, namely channel 80, by the electronics responsible for the bolometers' signal and then used during the offline analyses to tag the events generated by the injection of external light into the LDs. The green line identifies the bias for the light source, while the orange ones are the connections via optical fibres from the light source, through the 10/90 and 50/50 splitters, toward the detectors. Finally, the red line represents the PMT signal to be sampled by the digitiser. }\label{FIG:7}
\end{figure}

Inside the cryostat, at the 300 K flange, two 50/125 \(\mu\)m multimode optical fibres guide the generated light from outside and travel across the cooler stages, while thermalising along the way, until they reach the coldest stage, kept at around 15 mK. This preserves the structural integrity of the fibre cores during thermal cycles. Stage-by-stage thermal anchoring intercepts heat conducted along the fibres at intermediate cryostat stages, where the available cooling power is greater, thereby reducing the thermal load on the coldest stage. In addition to thermal anchoring, the optical injection line should be designed to minimise the transmission of unwanted infrared radiation to the light detectors, like its own blackbody radiation. At the tower level, inside the experimental volume, a couple of custom-made copper-PTFE connectors join each one of the two transporting fibres to a transparent polymethyl methacrylate optical fibre. The two transparent fibres run vertically along the tower, one per side, and their cladding is etched to induce spot-like emission at the height of each LD. The performed cuts are a few hundred \(\mu\)m deep and wide, with a 5 cm spacing among them. Localised emission improves illumination efficiency, reducing the source power required for a given LD signal and limiting residual optical power dissipated elsewhere in the experimental volume.
Everything else, the light source and monitoring devices, is kept outside the cryostat, at room temperature, leaving them more easily accessible. Light transport at room temperature is carried out with 50/125 \(\mu\)m multimode optical fibres. For the light source, a Thorlabs LPS-635-FC single-mode fibre-pigtailed laser diode was selected, with a central wavelength of about 635 nm. The source is fixed in position inside a screw-closed aluminium box, to make it easier to handle, with an input for the laser diode bias and a fibre output for the light. The output is then connected to a 1:2 optical splitter with a 90/10 split ratio and a further 1:2 optical splitter with a 50/50 split ratio following the 90\% output branch. These splitting chains nominally provides two balanced 45-45 optical outputs for detector light injection and a lower-power one for monitoring purposes. The first two are then connected to the 300 K flange of the cryostat with two distinct vacuum-compatible FC-PC connectors, while the third one is terminated in a room-temperature chamber with a 1-inch photomultiplier tube R6095 by Hamamatsu. The photomultiplier tube is biased at 650 V and its signal is sampled with a DT5724 CAEN digitiser. Additionally, a PT1000 thermometer, sampled by an NI6008 USB from National Instruments, is deployed to monitor the environmental temperature around the laser diode to cross-check for any instabilities, and a Tektronix AFG1062 arbitrary function generator works as a pulser for the laser diode's bias circuit (fig.\ref{FIG:7}). The handling of the pulser and the data acquisition of both monitoring devices, the thermometer and photomultiplier, is performed by custom-made software developed for the Optical Injection System.
\subsection{OIS: studies and applications}
\begin{figure}[]
  \centering
    \includegraphics[width=1.\columnwidth]{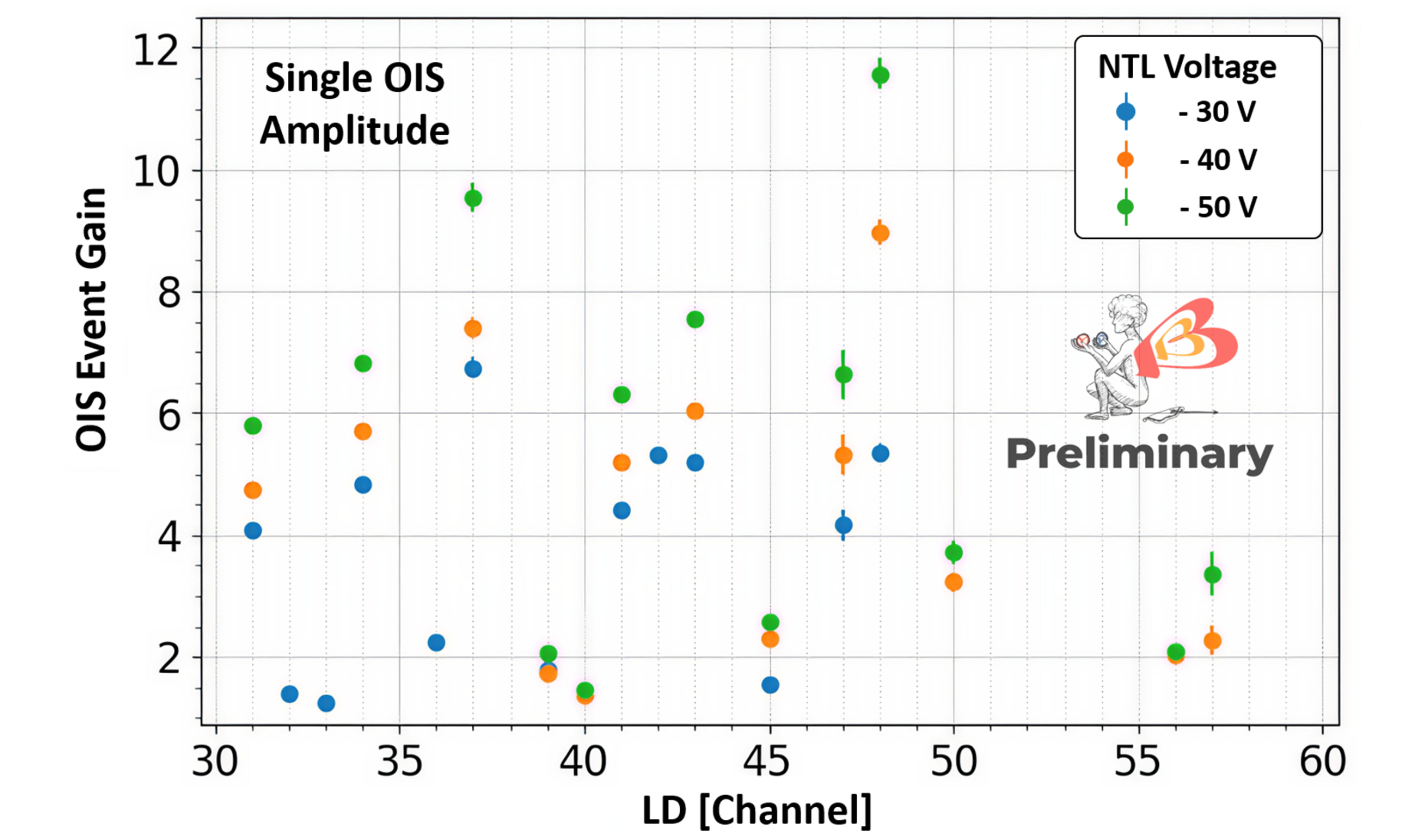}
    \caption{NTL gain characterisation for a sub group of channels which has been performed by injecting a fixed optical energy per pulse with the OIS light source. A few channels reported a higher standard deviation for the observed signals' amplitude, making their error bars more visible. The amplitude gain is reported in units of amplitude for the corresponding pulse without amplification.}\label{FIG:8}
\end{figure}

\begin{figure}
  \centering
    \includegraphics[width=1\columnwidth]{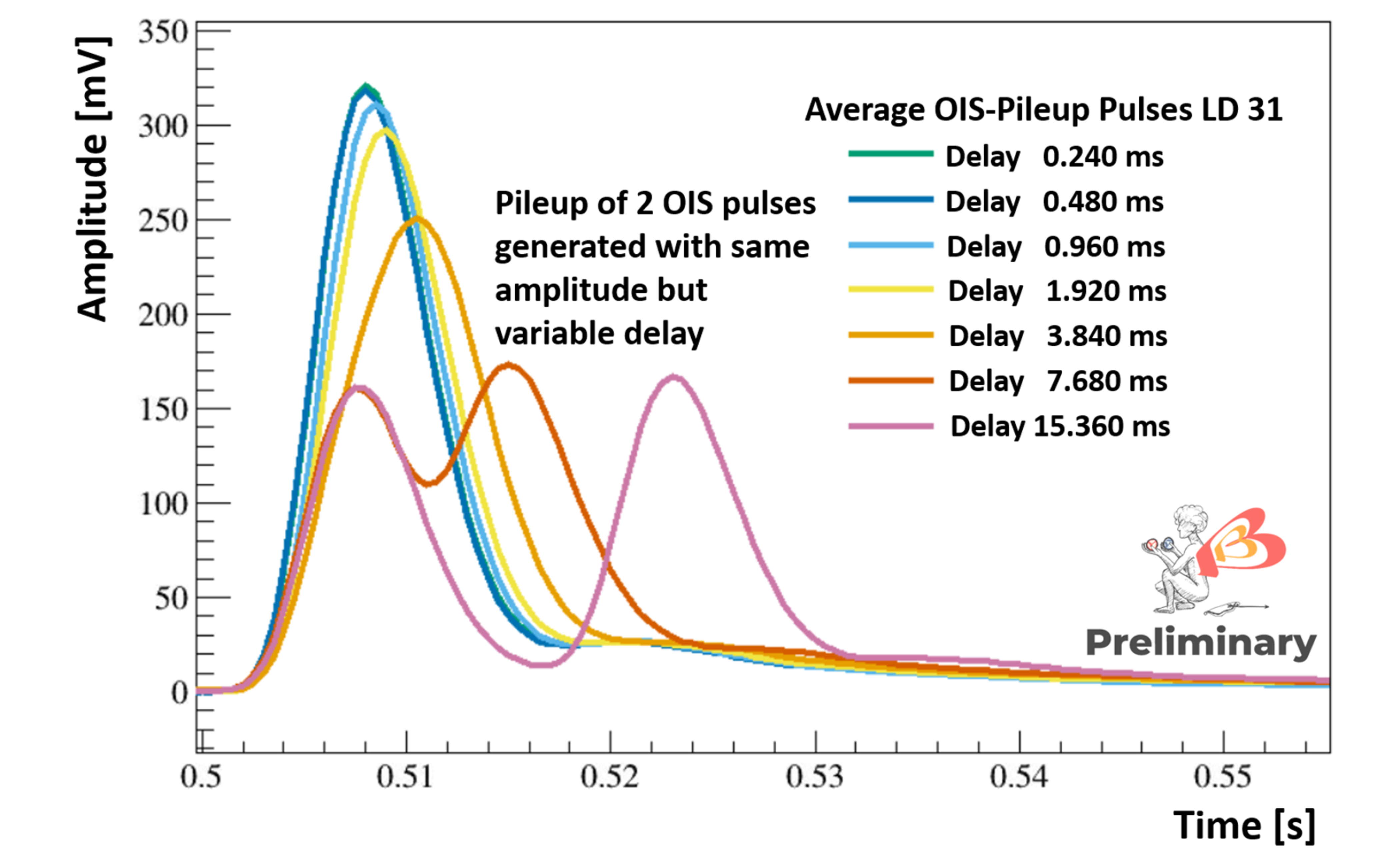}
    \caption{Average responses of LD-31 to pairs of OIS light pulses with equal injected energies (\(\alpha=0.5\)) and different programmed time separations, as indicated in the legend. The detector was operated at an NTL electrode bias of 0 V and sampled at 2 kHz. The reported delays represent the programmed separation between the optical pulses, rather than a measured detector resolving time.}\label{FIG:9}
\end{figure}

An apparatus like the OIS makes it possible to perform a wide range of tests, from LD and NTL-performance studies to physics event simulation. By fixing the amount of energy injected into our LDs, by holding constant both power and pulse duration, we first perform a study of the achievable gain thanks to NTL amplification, at different voltage biases, (fig.\ref{FIG:8}).
By firing a couple of pulses with the OIS, it is possible to simulate a physics pile-up event on the LDs. We define \(\alpha\) as the relative energy of the first pulse with respect to the sum of the two and \(\Delta t\) as the delay time between the two pulses. By fixing the total amount of injected energy, i.e. the sum of the energy from the two pulses, one can vary both \(\alpha\) and \(\Delta t\) to probe the difference in response of the LDs, allowing one to measure the pile-up rejection power in different conditions, (fig.\ref{FIG:9}). 
\section{Conclusions}
The VSTT serves as the second prototype tower for CUPID, and also the first full-scale CUPID prototype tower combining NTL-assisted light detectors with an Optical Injection System. The VSTT measurements demonstrated an improvement in the signal-to-noise ratio of approximately one order of magnitude compared with operation without NTL amplification. At the same time, the few tests carried out with the OIS have shown the potential of such a system to generate physics-like events for different purposes. Particularly, the possibility of generating pile-up-like events allows us to study the rejection techniques for our LDs and their performance at different energies, delay times, and relative amplitudes. The VSTT has recently concluded its data-taking campaign, and the analysis is currently ongoing. Further R\(\&\)D tests are currently planned for the near future, with the aim of both improving the performance of NTL-assisted LDs and conducting OIS-dedicated measurements.  

\section*{Acknowledgements}
The CUPID Collaboration thanks the directors and staff of the Laboratori Nazionali del Gran Sasso and the technical staff of our laboratories. This work was supported by the Istituto Nazionale di Fisica Nucleare (INFN); by the European Research Council (ERC) under the European Union Horizon 2020 program (H2020/2014-2020) with the ERC Advanced Grant no. 742345 (ERC-2016-ADG, project CROSS); by the Italian Ministry of University and Research (MIUR) through the grant Progetti di ricerca di Rilevante Interesse Nazionale (PRIN)  grant nos. 2017FJZMCJ, 2020H5L338, and 202273P73A; by the US National Science Foundation under grant nos. NSF-PHY-1401832, NSF-PHY-1614611, NSF-PHY-2412377, and NSF-PHY-1913374; by the French Agence Nationale de la Recherche (ANR) through the ANR-21-CE31-0014- CUPID-1. This material is also based upon work supported by the US Department of Energy (DOE) Oﬃce of Science under Contract Nos. DE-AC02-05CH11231 and DE-AC02-06CH11357; and by the DOE Oﬃce of Science, Oﬃce of Nuclear Physics under Contract Nos. DE-FG02-08ER41551, DE-SC0011091, DE-SC0012654, DE-SC0019316, DE-SC0019368, and DE-SC0020423. This work was also supported by the Russian Science Foundation under grant No. 18-12-00003. This research used resources of the National Energy Research Scientific Computing Center (NERSC). This work makes use of both the DIANA data analysis and APOLLO data acquisition software packages, which were developed by the CUORICINO, CUORE, LUCIFER and CUPID-0 Collaborations.












\bibliographystyle{model1-num-names}

\bibliography{myBibliography}
\bio{}
\end{document}